\documentclass[conference,10pt]{IEEEtran}
\IEEEoverridecommandlockouts
\usepackage{mathtools}
\usepackage{graphics, theorem, times, amsfonts, graphicx, amssymb, cite}
\usepackage{tikz}
\usetikzlibrary{shapes,arrows}
\usepackage{pgfplots}
\usepackage{color}
\usepackage{setspace}
\usepackage{rotating}
\usepackage{comment}
\usepackage[keeplastbox]{flushend}
\usepackage[font=small]{caption}
\usepackage{subcaption}
\usepackage[affil-it]{authblk}
\usepackage{bm}
\usepackage{algorithm}
\usepackage{algorithmic}
\usepackage{enumerate}
\usepackage{cite}

\makeatletter
\newcommand\fs@betterruled{%
  \def\@fs@cfont{\bfseries}\let\@fs@capt\floatc@ruled
  \def\@fs@pre{\vspace*{5pt}\hrule height.8pt depth0pt \kern2pt}%
  \def\@fs@post{\kern2pt\hrule\relax}%
  \def\@fs@mid{\kern2pt\hrule\kern2pt}%
  \let\@fs@iftopcapt\iftrue}
\floatstyle{betterruled}
\restylefloat{algorithm}
\makeatother

\input{mysymbol.sty}
\usepackage{needspace}

\usepackage{theorem}

\newtheorem{remark}{\hspace{0pt}\bf Remark}

\date{\today}

\def\E{\mathbb{E}}

\title{Wireless Power Control via Counterfactual Optimization of Graph Neural Networks}
\author{Navid Naderializadeh$^{*}$ \quad Mark Eisen$^{*}$ \quad Alejandro Ribeiro$^{\dagger}$ \thanks{{Supported by the Intel Science and Technology Center for Wireless Autonomous Systems. The authors are with $^{*}$Intel Labs and $^{\dagger}$University of Pennsylvania. E-mails: navid.naderializadeh@intel.com, mark.eisen@intel.com, aribeiro@seas.upenn.edu}.}}
\date{January 2020}

\begin{document}

\maketitle

\begin{abstract}
We consider the problem of downlink power control in wireless networks, consisting of multiple transmitter-receiver pairs communicating with each other over a single shared wireless medium. To mitigate the interference among concurrent transmissions, we leverage the network topology to create a graph neural network architecture, and we then use an unsupervised primal-dual counterfactual optimization approach to learn optimal power allocation decisions. We show how the counterfactual optimization technique allows us to guarantee a minimum rate constraint, which adapts to the network size, hence achieving the right balance between average and $5^{th}$ percentile user rates throughout a range of network configurations.
\end{abstract}

\begin{IEEEkeywords}
Wireless power control, graph neural networks, counterfactual optimization, primal-dual learning \end{IEEEkeywords}

\section{Introduction}
With the proliferation of ubiquitous wireless devices and services, wireless communication networks are becoming increasingly complex. In particular, the arrival of $5^{th}$ generation mobile networks (5G) will provide connectivity to devices ranging from sensors and cell phones to vehicles and drones, shifting the paradigm of how \emph{things} connect together. This will give rise to ultra-dense deployment scenarios, where a massive number of transmissions compete to obtain access to a limited amount of wireless resources.


One of the main drivers of higher throughput in 5G networks and beyond is leveraging the bandwidth that is available at higher frequency bands, such as the mmWave band. However, given the fact that the physical wireless resources are limited in nature, another way to enhance the performance of wireless networks is to improve the spectral efficiency. 
This becomes extremely challenging as the networks become denser, since the interference among concurrent transmissions can significantly hurt the network performance.

To deal with these challenges, there has been a plethora of work on \emph{radio resource management} in wireless networks. The approaches proposed in the literature use a wide variety of techniques in optimization, information, and game theories in order to attack various radio resource management sub-problems, including power control, link scheduling, cell association, sub-carrier assignment, and beamforming~\cite{madan2010cell, shi2011iteratively, yu2013multicell, wu2013flashlinq, naderializadeh2014itlinq, yi2015itlinq+, song2016game, shen2017fplinq}.

Nevertheless, solving the radio resource management problem in its most general form is NP-hard, implying that as the network size increases, it becomes more challenging to derive an optimal solution~\cite{luo2008dynamic,liu2013complexity}. That is why most prior works in the literature devise approximate solutions in various regimes of system parameters. With the recent success of machine learning, and particularly deep learning, over the past few years, learning-based algorithms have also been shown to result in promising solutions for resource management in wireless networks~\cite{lee2019deep,liang2019deep,eisen2019learning,naderializadeh2019multiple}. More recently, the natural graph structure of wireless interference patterns has been leveraged in graph neural network (GNN) architectures \cite{eisen2019optimal,lee2019graph, shen2019graph} that are more suited to scalability and transference.

In this paper, we consider a wireless interference channel comprising multiple transmitter-receiver pairs, and seek a power control policy that mitigates the interference among concurrent transmissions with respect to both overall system performance and fairness across pairs. We model the network topology by a conflict graph, where each edge represents the interference between pairs that are strong compared to the signal power levels, while the absent edges correspond to interference links that are weak enough to be treated as noise~\cite{geng2015optimality}. We then leverage the instantaneous conflict graph in a GNN that outputs a power allocation decision for each transmitter. We pose the power control problem as the  optimization of the filter weights of the GNN such that a network-wide convex utility function is maximized subject to some minimum rate constraints for all receivers.

Channel values in wireless networks fluctuate from time to time and from topology to topology. Therefore, even for a given density of transmitters and receivers, a fixed and strict minimum rate constraint may not be satisfiable for some of the receivers with poor channel conditions and is hard to define a priori. Hence, we introduce a counterfactual optimization formulation, in which an adaptive slack variable is subtracted from the minimum rate constraints \cite{chamon2020counterfactual}. We then utilize a primal-dual optimization algorithm to learn optimal policies and their associated optimal constraint slacks. We demonstrate through simulation results how our proposed framework learns a power control strategy that strikes a balance between the sum-rate and cell-edge performance---quantified by the $5^{th}$ percentile rate achieved by the users. In addition, we illustrate how the algorithm adaptively tunes the slack variable, hence the minimum rate constraints for the receivers, given the density of the network. 

The rest of this paper is organized as follows. In Section~\ref{sec_problem_formulation}, we present the system model and formulate the problem. In Section~\ref{sec:REGNN}, we provide the details of the GNN architecture. In Section~\ref{sec:CF}, we show how counterfactual optimization adapts the constraints as needed. In Section~\ref{sec:sim}, we present our simulation results. Finally, we conclude the paper in Section~\ref{sec:conc}.

\section{System Model and Problem Formulation} \label{sec_problem_formulation}
We consider a wireless interference network with a set of $m$ transmitters $\{\mathsf{Tx}_i\}_{i=1}^m$ and a set of $m$ receivers $\{\mathsf{Rx}_j\}_{j=1}^m$, where each transmitter ${\mathsf{Tx}_i}$ intends to communicate to its corresponding receiver ${\mathsf{Rx}_i}$. The channel gain between each transmitter $\mathsf{Tx}_i$ and each receiver $\mathsf{Rx}_j$ in the network is a random variable denoted by $h_{ij}$. We collect all the channel gains across the network in a square matrix, denoted by $\bbH \in \ccalH \subseteq \mathbb{C}^{m \times m}$. Each channel gain in $\bbH$ is composed of a constant long-term component, resulting from path loss and shadowing---due to signal attenuation from the physical distance between the transmitter and receiver nodes, alongside deviations thanks to obstacles in the environment---and a short-term fast fading component---a result of multi-path propagation in the channel and node mobility. In general, we assume that $\bbH$ is drawn from a joint probability distribution~$f(\bbH)$.

Assuming that all transmissions occur at the same time and on the same frequency band, they will cause interference on each other. Therefore, it is imperative for each transmitter to set its transmit power such that a global network-wide objective function is optimized. In particular, for each channel realization $\bbH$, we denote the vector of power allocation variables by $\bbp\in\reals^m$, whose $i^{th}$ component, $p_i$, represents the transmit power allocated to transmitter $\mathsf{Tx}_i$. This implies that the signal-to-interference-plus-noise ratio (SINR) at each receiver $\mathsf{Rx}_i$ can be written as
\begin{align}\label{eq:SINR_def}
\mathsf{SINR}_i(\bbH, \bbp) = \frac{ |h_{ii}|^2 p_i(\bbH)} {\sigma^2 + \sum_{j \neq i} |h_{ji}|^2 p_j(\bbH)},\quad \forall i \in [m],
\end{align}
where $\sigma^2$ denotes the noise variance, and $[m]$ is defined as $[m]\triangleq\{1,...,m\}$. The Shannon capacity of the link between transmitter $\mathsf{Tx}_i$ and receiver $\mathsf{Rx}_i$ is then given by
\begin{align}
C_i(\bbH, \bbp) = \log_2(1 + \mathsf{SINR}_i(\bbH, \bbp)).
\end{align}

Due to the aforementioned short-term fading phenomenon, channel realizations vary over time, implying that the power allocation variables also need to be modified temporally. This motivates considering the ergodic average~$x_i = \E [C_i(\bbH)] \in \reals$, to capture the throughput experienced by each receiver over a long period of time. The goal is to determine a power allocation policy $\bbphi(\bbH, \bbtheta)$ parameterized by a fixed parameter vector $\bbtheta \in \reals^q$, where, for each channel realization $\bbH$, the transmit powers are determined by $\bbp = \bbphi(\bbH,\bbtheta)$. We formulate the power allocation problem to find the parameter vector $\bbtheta$ that provides the best performing policy, i.e.,
\begin{alignat}{3} \label{eq_param_problem}
    &  \max_{\bbtheta,\bbx} ~  && \mathcal{U}(\bbx),             \\
        &  \st                     && x_i       \leq  \E_{\bbH} \left[ C_i(\bbH, \bbphi(\bbH,\bbtheta)) \right],~ \forall i\in[m],   \nonumber \\
        &   \ && x_i \geq C_{i,\min},~ \forall i\in[m], \nonumber\\
        &   \ && \bbphi(\bbH,\bbtheta) \in  [0,P_{\max}]^m.   \nonumber%
\end{alignat}
In the above optimization problem, $\mathcal{U}(\bbx)$ denotes a convex function of the receivers' ergodic rates throughout the network, $C_{i,\min}$ denotes a minimum capacity that each receiver needs to satisfy, and $P_{\max}$ denotes the maximum transmit power. The minimum capacity constraints are included so as to avoid allocating all resources to ``cell-center'' receivers, hence balancing the power control policy to treat ``cell-center'' and ``cell-edge'' receivers \emph{fairly}.

The problem in~\eqref{eq_param_problem} is generally challenging to solve, mainly due to the non-convexity of the constraints. Moreover, aside from the effort in solving \eqref{eq_param_problem}, the choice of parameterization function $\bbphi$ is critical in achieving an optimal policy with good practical performance. Fully-connected deep neural networks (DNNs) are a proper choice here, due to their universality property, which states that given enough depth and/or width,  
they have sufficient expressive power to approximate any function with any desired accuracy~\cite{hornik1989multilayer,eisen2019learning}. However, despite their theoretical properties, such a parameterization does not scale well---as the parameter dimension $q$ grows with number of transmitter-receiver pairs $m$ in the network---and more critically does not generalize over varying network topologies. In the next section, we discuss and develop a graph neural network architecture suitable for solving the power allocation problem in networks of any size.

\section{Random Edge Graph Neural Networks}\label{sec:REGNN}
We present the \emph{random edge graph neural network (REGNN)} architecture as a parameterization for the resource management policy. Broadly speaking, graph neural networks (GNNs) can be viewed as a generalization of convolutional neural network (CNN) architectures, whose popularity and practical benefits stem largely from their significantly reduced parameter dimension relative to traditional DNNs, their invariance to input size, and their so-called translation equivariance.

Graph neural networks generalize the convolutional operations performed in CNNs with a convolution performed on arbitrarily structured data \cite{henaff2015deep}. This structure is given in the form of a graph $\ccalG = (\ccalV, \ccalE)$, where $\ccalV := [m]$ are the nodes of the graph connected by weighted edges $\ccalE$. We further use the matrix $\bbS \in \reals_{+}^{m \times m}$ as a graph shift operator, that encodes the weights of edges $\ccalE$. The elements $S_{ij}$ take on higher values when node $i$ is closely related to node $j$, smaller values when they are less related, and a value of 0 if they are unrelated. The graph convolution of input signal $\bby \in \reals^m$---whose $i^{th}$ element $y_i$ is the signal value at transmitter $\mathsf{Tx}_i$---and filter $\bbalpha \in \reals^K$ with respect to the graph encoded in $\bbS$ is a vector $\bbz \in \reals^{m}$, whose $j^{th}$ component is defined as
\begin{align}\label{eq_graph_conv}
z_j := [\bbalpha *_{\bbS} \bbx]_j := \sum_{k=0}^K \alpha_k [\bbS^k \bby]_j.
\end{align}
Observe that the term $\bbS^k$ shifts the elements of $\bbx$ in $k$ turns according to the weights and structure defined in $\bbS$. 

A GNN is constructed with a sequence of $L$ so-called hidden layers, where the output of layer $l-1$ is fed as an input to layer $l$. Denote by $\bby_l$ the input to layer $l$, and by $\bbalpha_l$ the graph filter at layer $l$. With shift operator $\bbS$, the output of layer $l$, denoted by $\bby_{l+1}$, is computed as a composition of the graph filter $\bbalpha_l$ and a pointwise, nonlinear function $\bbsigma_l(\cdot)$, i.e.,
\begin{align}\label{eq_graph_layer}
\bby_{l+1} := \bbsigma_l(\bbalpha_l *_{\bbS} \bby_l).
\end{align}
The full GNN is then formed as the composition of layer operations as in \eqref{eq_graph_layer} for $l\in[L]$. The input to the GNN is given as the initial graph input signal $\bby_0 \in \reals^m$, defined on the nodes $\ccalV$. While standard applications feature a fixed graph, we may also consider more generically an input graph $\bbS \in \reals^{m \times m}_+$---i.e., an input signal on the edges $\ccalE$. When such inputs are drawn randomly from some distribution, this may otherwise be considered as a graph with random edge weights.

In the wireless interference network defined in Section~\ref{sec_problem_formulation}, a graph can be readily formed using the transmitter-receiver channel gains contained in the channel matrix $\bbH$. We define the graph $\bbS := \bbg(\bbH)$, where $\bbg: \reals_+^{m \times m} \rightarrow \reals_+^{m \times m}$ is some function that preserves the sparsity pattern and node ordering of the channel matrix $\bbH$. Simple choices for $\bbg(\cdot)$ may include element-wise magnitude $[\bbg(\bbH)]_{ij} := |h_{ij}|$ or squared magnitude $[\bbg(\bbH)]_{ij} := |h_{ij}|^2$. In this work, we use the information-theoretic optimality condition for treating interference as noise, derived in~\cite{geng2015optimality}, to classify the interference links between all non-associated transmitter-receiver pairs as strong or weak. In particular, we take an approach similar to~\cite{naderializadeh2014itlinq}, where for each interference link between $\mathsf{Tx}_i$ and $\mathsf{Rx}_j, j\neq i$, we define indicator variables
\begin{align}\label{eq:TIN_graph_indicator}
I_{ij} = 1 \text{ iff } \tfrac{P_{\max}|h_{ij}|^2}{\sigma^2} \geq M \left(\tfrac{P_{\max}\min\{|h_{ii}|^2, |h_{jj}|^2\}}{\sigma^2}\right)^\eta,
\end{align}
where $M$ and $\eta$ are design parameters, controlling the sparsity of the graph. We then define $\bbg(\cdot)$ as $[\bbg(\bbH)]_{ij} := I_{ij}|h_{ij}|^2$, where for each direct link between $\mathsf{Tx}_i$ and $\mathsf{Rx}_i$, we set $I_{ii}=1$. We finally normalize the resulting matrix $\bbS$ by its $2-$norm.

As the edge weights of $\bbS$ are derived from the random channel gain values, we consider the previously described case of GNNs with random input graphs---called the random edge graph neural network (REGNN)---with edges drawn from joint distribution $f(\bbg(\bbH))$ \cite{eisen2019optimal} . The full REGNN parameterization $\bbphi(\bbH,\bbtheta)$ of the resource management policy can then be described as a GNN with a constant input $\bby_0 := \mathbf{1}$, i.e.,
\begin{align}\label{eq_gnn_param}
&\bbphi(\bbH,\bbtheta) :=\bbsigma_L(\bbalpha_L *_{\bbg(\bbH)} (\hdots( \bbsigma_1(\bbalpha_1 *_{\bbg(\bbH)} \bby_0)\hdots))).
\end{align}
where the parameter $\bbtheta$ contains the $L$ sets of filter weights, i.e., $\bbtheta = \{\bbalpha_l\}_{l=1}^L$. The final nonlinear activation $\bbsigma_L$ can be chosen to scale the output between $[0, P_{\max}]$. Note that with a filter length of $K_l$ at the $l^{th}$ layer, the total number of parameters in a GNN is $q = \sum_{l=1}^L K_l$, a number significantly smaller than that of a fully connected DNN and invariant to the size of the input graph, i.e., number of transmitter-receiver pairs.

We point out that a key feature of REGNNs that make them well suited for learning in wireless networks lies in a structural property called \emph{permutation equivariance}. Permutation equivariance implies that any permutation of the rows and columns of the channel matrix $\bbH$---i.e., a  relabeling of the indices of transmitter-receiver pairs in the wireless network---will result in an equally permuted output for any REGNN $\bbphi(\bbH,\bbtheta)$ as defined in \eqref{eq_gnn_param}---see \cite{eisen2019optimal}. This property is valuable in wireless networks because it can facilitate the training of a REGNN to operate over many different geometric configurations of the transmitters and receivers in the network, which will invariably change over time in practice. We will demonstrate the effectiveness in using REGNNs to achieve strong performance over a wide range of network configurations in Section~\ref{sec:sim}.

\section{Counterfactual Optimization}\label{sec:CF}

While we may utilize the REGNN to parameterize the resource allocation policy, a training algorithm must be used to find the proper set of GNN filter weights $\bbtheta = \{\bbalpha_l\}_{l=1}^L$ that performs well under the metrics and constraints defined in \eqref{eq_param_problem}. Training the filter weights here is not straightforward in that the resulting policy must not only maximize the utility function $\ccalU(\bbx)$, but also satisfy the minimum rate constraints $x_i \geq C_{i,\min}$. While constraints can generally be satisfied with a Lagrangian dual function, this requires explicit a priori knowledge of the minimum achievable rate $C_{i,\min}$. However, this is generally not known in practice, as complex interference patterns between concurrent transmissions in different network densities may make some lower bounds infeasible.

We address this problem with what may be referred to as a \emph{counterfactual} \cite{chamon2020counterfactual}. That is, we consider a slack term $s_i$ for the $i^{th}$ minimum capacity constraint, and try to find the optimal policy under the loosened constraint. Any increase in slack $s_i$ will render a solution further from the intended solution of \eqref{eq_param_problem}; as such, we further seek to minimize $s_i$ under the condition that the problem remains feasible. More formally, we augment \eqref{eq_param_problem} with the counterfactual slack variable $\bbs \in \reals^m_{+}$ as
\begin{alignat}{3} \label{eq_slack_problem}
    &  \max_{\bbtheta,\bbx, \bbs} ~  && \mathcal{U}(\bbx) - \frac{1}{2}\|\bbs\|^2,             \\
        &  \st                     && x_i       \leq   \E_{\bbH} \left[ C_i(\bbH, \bbphi(\bbH,\bbtheta)) \right],~ \forall i\in[m],   \nonumber \\
        &   \ && x_i \geq C_{i,\min} - s_i,~ \forall i\in[m], \nonumber\\
        &   \ && \bbphi(\bbH,\bbtheta) \in  [0,P_{\max}]^m, \quad \bbs \geq \bb0.   \nonumber%
\end{alignat}
%
In \eqref{eq_slack_problem}, along with optimizing the REGNN parameters $\bbtheta$ and ergodic average rates $\bbx$, we also minimize the value of the slack $\bbs$ that makes the problem feasible. Increasing $\bbs$ will lessen the achieved objective value in \eqref{eq_slack_problem}. However, too small a slack may make the constraints too tight to satisfy, rendering the problem unsolvable. The value in the counterfactual formulation lies in the fact that, should the preferred $C_{i,\min}$ be unrealizable, the optimization of slack variables will implicitly loosen this requirement just enough to find a solution.

We proceed to derive the training algorithm by introducing the Lagrangian function, with non-negative dual multipliers $\bblambda, \bbmu \in \reals^{m}_+$ associated with each constraint in \eqref{eq_slack_problem}, as
\begin{align}\label{eq_param_lagrangian}
   \ccalL(\bbtheta,&\bbx,\bbs, \bblambda,\mu) := \mathcal{U}(\bbx) - \frac{1}{2}\|\bbs\|^2
   \\
   & - \bblambda^T\left[\bbx -     \E_{\bbH} \bbC(\bbH, \bbphi(\bbH,\bbtheta)) \right] - \bbmu^T \left[ \bbC_{\min} - \bbs - \bbx\right].  \nonumber
\end{align}
The Lagrangian in \eqref{eq_param_lagrangian} provides a single, unconstrained objective function, which we can optimize using gradient-based methods. In particular, we seek to maximize over the so-called primal variables $\bbtheta, \bbx, \bbs$, while subsequently minimizing over the dual variables $\bblambda, \bbmu$, i.e.
\begin{align}\label{eq_dual_problem}
\min_{\bblambda, \bbmu \geq \bb0} \max_{\bbtheta, \bbx, \bbs} \ccalL(\bbtheta,\bbx,\bbs,\bblambda,\bbmu).
\end{align}
We can now define the updates over an iteration index $k$ for each primal and dual variable by either adding or subtracting the partial gradient of $\ccalL(\bbtheta,\bbx,\bbs,\bblambda,\bbmu)$ with respect to that variable. For the primal variables, this gives us the updates,
\begin{align}
\bbtheta_{k+1} &= \bbtheta_k + \gamma_1 \nabla_{\bbtheta}
\E_{\bbH} \bbC(\bbH, \bbphi(\bbH,\bbtheta_k))\bblambda_k , \label{eq_pd_update1} \\
\bbx_{k+1} &=\bbx_k  +  \gamma_{2} ( \nabla \ccalU(\bbx_k)  - \bblambda_k + \bbmu_k) , \label{eq_pd_update2} \\ 
\bbs_{k+1} &= \left[\bbs_k  +  \gamma_{3} ( \bbmu_k - \bbs_k) \right]_+, \label{eq_pd_update3}
\end{align}
where $\gamma_{1} ,\gamma_{2}, \gamma_{3} >0$ denote learning rates corresponding to the primal variables. Note that in addition to updating $\bbtheta_k$ and $\bbx_k$ in \eqref{eq_pd_update1}-\eqref{eq_pd_update2}, the counterfactual formulation updates the slack variable $\bbs_k$ as the difference between the current slack and dual variables. Likewise, we descend on the dual variables using the associated partial gradients of the Lagrangian, i.e.,
\begin{align}
\bblambda_{k+1} &=  \bblambda_k - \gamma_{4} \left( \E_{\bbH} \bbC(\bbH,\bbphi(\bbH,\bbtheta_{k})) -\bbx_{k} \right),\label{eq_pd_update4} \\
\bbmu_{k+1} &= \left[ \bbmu_k - \gamma_{5} \left( \bbx_k + \bbs_k - \bbC_{\min} \right) \right]_+,\label{eq_pd_update5}
\end{align}
with $\gamma_{4}, \gamma_5 >0$ representing learning rates corresponding to the dual variables. The primal-dual gradient updates in \eqref{eq_pd_update1}-\eqref{eq_pd_update5} successively move the primal and dual variables towards maximum and minimum points of the Lagrangian dual function, respectively. The complete counterfactual primal-dual learning algorithm is summarized in Algorithm \ref{alg:learning}.

\begin{remark}
\normalfont
Our proposed method is \emph{unsupervised} in the sense that we train the REGNN weights to optimize the utility and constraints in~\eqref{eq_slack_problem} directly rather than with labeled solutions. Therefore, this algorithm can be applied to all different types of radio resource management problems, whose objectives and constraints can be formulated as in~\eqref{eq_slack_problem}, without the need to have any optimal solutions beforehand.
\end{remark}

\begin{remark}
\normalfont
We point out that evaluating the updates in \eqref{eq_pd_update1}-\eqref{eq_pd_update5} may require computing potentially challenging gradients and expectations. The gradients in these updates can be replaced with well-known \emph{model-free} gradient estimation methods that can be obtained with function evaluations and channel sampling---see \cite{eisen2019learning} for details on these approaches.
\end{remark}

\section{Simulation Results}\label{sec:sim}
\setlength{\textfloatsep}{0pt}
\begin{algorithm}[t]
\begin{algorithmic}[1]
\STATE \textbf{Parameters:} REGNN model (e.g., filter lengths $\{K_l\}_{l=1}^L$)\vspace{-.17in}
\STATE \textbf{Input:} Initial values $\bbtheta_0, \bbx_0, \bblambda_0, \bbmu_0, \bbs_0 = \bb0$
\FOR{$k = 0,1,2,\hdots$}
     \STATE Update primal variables [\text{cf.~}\eqref{eq_pd_update1}-\eqref{eq_pd_update2}]\vspace{-.05in}
         \begin{align*}
	\bbtheta_{k+1} &= \bbtheta_k + \gamma_1 \nabla_{\bbtheta}
\E_{\bbH} \bbC(\bbH, \bbphi(\bbH,\bbtheta_k))\bblambda_k , \\
	\bbx_{k+1} &=\bbx_k  +  \gamma_{2} ( \nabla \ccalU(\bbx_k)  - \bblambda_k + \bbmu_k).
	\end{align*}\vspace{-.17in}
	\STATE Update slack variable [\text{cf.~}\eqref{eq_pd_update3}]\vspace{-.05in}
         \begin{align*}
	\bbs_{k+1} &= \left[\bbs_k  +  \gamma_{3} ( \bbmu_k - \bbs_k) \right]_+.
	\end{align*}\vspace{-.17in}
	\STATE Update dual variables [\text{cf.~}\eqref{eq_pd_update4}-\eqref{eq_pd_update5}]\vspace{-.05in}
         \begin{align*}
	\bblambda_{k+1} &= \bblambda_k - \gamma_{4} \left( \E_{\bbH} \bbC(\bbH,\bbphi(\bbH,\bbtheta_{k})) -\bbx_{k} \right) , 	 \\
	\bbmu_{k+1} &= \left[ \bbmu_k - \gamma_{5} \left( \bbx_k + \bbs_k - \bbC_{\min} \right) \right]_+.
	\end{align*}
\ENDFOR

\end{algorithmic}
\caption{Counterfactual Primal-Dual Learning}\label{alg:learning}
\end{algorithm}

We consider wireless networks with $m\in\{6,8,10,12,14\}$ transmitter-receiver pairs, dropped randomly within a square area of side length 500m. We drop the transmitters uniformly at random within the network area, and ensure a minimum pairwise distance of 35m between them. Afterwards, for each transmitter, a receiver is dropped within an annulus centered at the transmitter, with inner and outer radii of 10m and 100m respectively, according to a skewed distribution that biases the receiver's location towards its serving transmitter. Each drop is then run for 200 steps. The long-term channel model consists of a standard dual-slope path-loss model~\cite{zhang2015downlink,andrews2016we} and log-normal shadowing with 7 dB standard deviation. We also model short-term Rayleigh fading using the sum of sinusoids (SoS) technique proposed in~\cite{li2002simulation}. The bandwidth is taken to be 10 MHz, the noise power spectral density is assumed to be -174 dBm/Hz, and the maximum transmit power is taken to be $P_{\max} = 10$ dBm. We utilize a sum-rate network utility function $\mathcal{U}(\bbx) = \sum_{i=1}^m x_i$, and we set the minimum capacity to $C_{i,\min} = 2$ bps/Hz for all receivers.

As for the learning parameters, we consider a GNN architecture with 3 hidden layers, each containing 4 features and a filter of size 4, and a ReLU activation function. We use $M=1$ and $\eta=0.6$ to build the underlying GNN graph as in~\eqref{eq:TIN_graph_indicator}. We consider a batch size of 200 consecutive samples within each drop. The learning rates for the primal, dual, and slack variables are set to $2\times10^{-2}$, $10^{-2}$, and $10^{-3}$, respectively. 
We consider a unique slack variable for all the minimum-rate constraints. We also restrict the power allocation decisions to be binary, i.e., each transmitter at any step either remains silent, or transmits with full power. We perform training on 4000 random and independent drops (realizations of transmitter/receiver locations), and we then conduct a final test on a set of 500 test drops.

\begin{figure}[t]
\centering
\includegraphics[trim = .1in .1in 0.1in .1in, clip,width=0.48\textwidth]{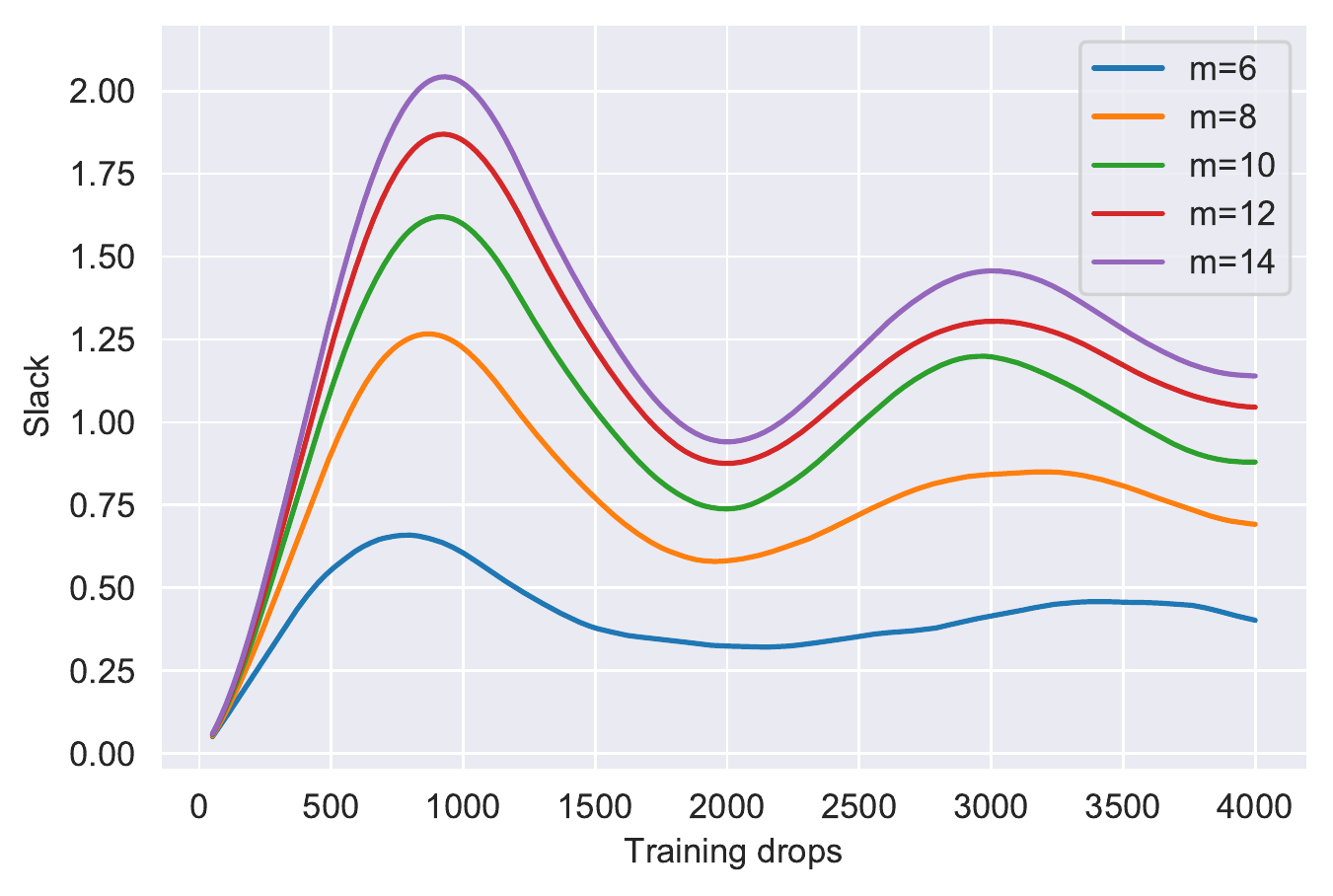}
\caption{Evolution of the slack variable during training for networks with $6-14$ transmitter-receiver pairs.}
\label{fig:slack}
\end{figure}

Figure~\ref{fig:slack} illustrates how the slack variable evolves during the course of training for different network densities. 
Increasing the number of transmitter-receiver pairs gives rise to higher levels of interference and lowers the average achievable rate of each receiver, making it harder to satisfy its minimum rate constraint. Therefore, as Figure~\ref{fig:slack} shows, our proposed algorithm indeed learns how to adaptively elevate the slack variable for denser deployments so as to make the optimization problem feasible and maximize the desired utility function.

\begin{figure}[t]
\centering
\includegraphics[trim = .1in .1in 0.1in .1in, clip,width=0.48\textwidth]{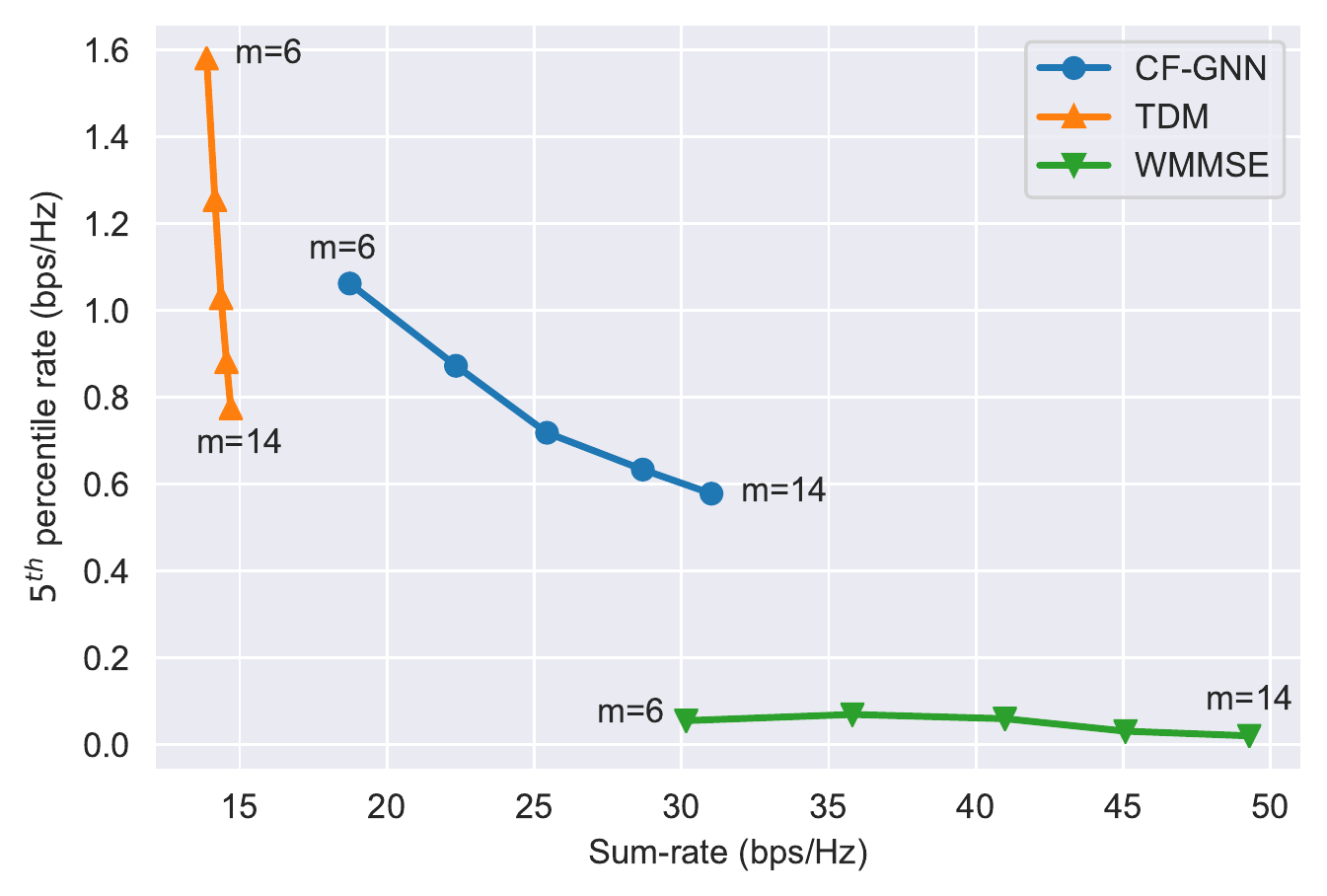}
\caption{The trade-off between achievable sum-rate and $5^{th}$ percentile rate by our proposed algorithm (CF-GNN) and the baseline algorithms for $6-14$ transmitter-receiver pairs.}
\label{fig:sumrate_vs_5perc}
\end{figure}

Moreover, Figure~\ref{fig:sumrate_vs_5perc} shows the achievable sum-rates and $5^{th}$ percentile rates of our proposed algorithm for different numbers of transmitter-receiver pairs, as compared with two baselines of time division multiplexing or \emph{TDM} (transmitters activated in a round-robin fashion), and weighted minimum mean-squared error or \emph{WMMSE}~\cite{shi2011iteratively}. As the figure shows, these two baselines represent two ends of a spectrum: TDM is completely fair across all pairs in the network, hence achieving excellent $5^{th}$ percentile rate performance, at the expense of poor sum-rate. WMMSE, on the other hand, merely optimizes sum-rate, hence sacrificing most of the pairs that are experiencing poor channel conditions. Our proposed method, however, demonstrates a superior trade-off between sum-rate and $5^{th}$ percentile rate, balancing the rates experienced by ``cell-center'' and ``cell-edge'' receivers. In particular, it achieves sum-rate gains of up to 110\% and $5^{th}$ percentile rate gains of up to 2740\% over TDM and WMMSE, respectively.

\section{Concluding Remarks}\label{sec:conc}
In this paper, we considered the problem of downlink power control in wireless networks with multiple transmitter-receiver pairs. We parametrized the power control policy as a graph neural network, whose edge weights are derived from the channel gains between the transmitters and receivers. We then proposed a primal-dual gradient-based optimization algorithm based on counterfactuals, which learns a power control policy that maximizes a convex network utility function with adaptive minimum rate constraints tuned to the actual network conditions. Simulation results show the superiority of our proposed algorithm compared to baseline methods in terms of the trade-off between average and $5^{th}$ percentile user rates.

\bibliographystyle{IEEEtran}
\bibliography{references}

\end{document}